\documentclass{emulateapj}

\journalinfo{\textsc{The Astrophysical Journal Letters}}

\begin{document}

\title{An IR--Selected Galaxy Cluster at $z=1.41$}

\author{S.A.\ Stanford\altaffilmark{1,}\altaffilmark{2}}
 
\author{Peter R.\ Eisenhardt\altaffilmark{3}} 

\author{Mark Brodwin\altaffilmark{3}}

\author{Anthony H.\ Gonzalez\altaffilmark{4}}
 
\author{Daniel Stern\altaffilmark{3}}

\author{Buell Jannuzi\altaffilmark{5}}

\author{Arjun Dey\altaffilmark{5}}

\author{Michael J.\ I.\ Brown\altaffilmark{6}}

\author{Eric McKenzie\altaffilmark{4}}

\author{Richard Elston\altaffilmark{4,}\altaffilmark{7}}

\altaffiltext{1}{University of California, Davis, CA 95616; adam@igpp.ucllnl.org}

\altaffiltext{2}{Institute of
Geophysics and Planetary Physics, Lawrence Livermore National
Laboratory, Livermore, CA 94551} 

\altaffiltext{3}{Jet Propulsion
Laboratory, California Institute of Technology, Pasadena, CA 91109; prme@kromos.jpl.nasa.gov; Mark.Brodwin@jpl.nasa.gov; stern@thisvi.jpl.nasa.gov}

\altaffiltext{4}{Department of
Astronomy, University of Florida, Gainesville, FL 32611; anthony@astro.ufl.edu; eric@astro.ufl.edu}

\altaffiltext{5}{National Optical Astronomy Observatories 
Tucson, AZ  85726-6732; jannuzi@noao.edu; dey@noao.edu}

\altaffiltext{6}{Department of Astrophysical Sciences, Peyton Hall, Princeton University, Princeton, NJ 08544; mbrown@astro.princeton.edu}

\altaffiltext{7}{Deceased}

\newpage

\shorttitle{IR GALAXY CLUSTER AT $z=1.4$}
\shortauthors{STANFORD ET AL.}

\begin{abstract} We report the discovery of a galaxy cluster at $z =
1.41$.  ISCS~J143809+341419 was found in the {\em Spitzer}/IRAC Shallow Survey of the Bo\"otes field
in the NOAO Deep Wide-Field Survey carried out by IRAC.  The cluster
candidate was initially identified as a high density region of objects
with photometric redshifts in the range $1.3 < z < 1.5$.  Optical
spectroscopy of a limited number of objects in the region shows that 5
galaxies within a $\sim$120 arcsec diameter region lie at $z = 1.41
\pm 0.01$.  Most of these member galaxies have broad--band colors
consistent with the expected spectral energy distribution of a
passively--evolving elliptical galaxy formed at high redshift. The
redshift of ISCS~J143809+341419 is the highest currently known for a
spectroscopically-confirmed cluster of galaxies.
    
\end{abstract}


\section{Introduction}

High-redshift galaxy clusters provide important tools in the study of
galaxy formation and evolution.  The galaxy populations of rich
cluster cores at $z > 1$ tend to be dominated by massive ellipticals
\citep[e.g.][]{cl1,cl3,0910,fp,1252acs,1252,acs,mullis,0910acs}, which are
useful probes of galaxy evolution because their stellar populations
appear to be relatively simple.  As the highest overdensity regions,
clusters should contain the oldest (and most massive) galaxies, so
that by probing $z > 1$ we can determine useful constraints on the
galaxy formation process. The connection between such stellar
populations and the ellipticals in present--epoch clusters
\citep{Bower} has been the subject of extensive analysis in recent
years \citep*[e.g.][]{SED2,pvd98,lubin,klf,dk00c}.  The nature of elliptical
galaxy formation in clusters at $z > 1$ depends strongly on the
importance and mode of merging in assembling the stellar mass
\citep{vdF01,cc05}.  Beyond $z\sim1$ in cosmologically-flat CDM
models, the amount of merging occurring within the prior $\sim$1 Gyr
is large and should seriously inflate the locus of early--type galaxy
colors due to interaction--induced starbursts \citep{cough},
although evidence exists for ``dry" mergers that could alleviate this
issue \citep{pvd05}.  On the other hand, models in which
ellipticals formed by a monolithic collapse at high-$z$  predict a tight color--magnitude
relation out to at least $z \sim 2$ for reasonable cosmologies \citep*{ELS}.  The
identification of clusters at $z > 1$ and the characterization of
their galaxy populations provides a powerful means of testing
elliptical galaxy formation theories.

One proven method of finding $z > 1$ cluster candidates is deep infrared
sky surveys.  Because they are massive and their rest-frame light peaks at $1-2 \mu$m,
early-type galaxies stand out from the background in the
observed-frame near- and mid-IR light out to $z \sim 2$.  Therefore,
to the extent that galaxy clusters are composed of galaxies, near- and
mid-IR imaging surveys offer a viable alternative to X--ray searches
for very high redshift clusters.  Both methods need to be exploited to
sample the entire range of galaxy clusters in the key $1 < z < 2$
epoch because it is unclear when and how massive galaxies and the
associated intracluster medium (ICM) are formed.  One advantage of using
infrared imaging surveys is that, in addition to identification of
massive clusters with an ICM that would be detected by an X-ray
survey, they can be extended to probe to lower mass scales and
identify the building blocks that yield massive clusters at lower
redshift.  We describe here the discovery of a $z > 1$ cluster which
resulted from an imaging survey conducted by the InfraRed Array Camera
\citep[IRAC;][]{IRAC} onboard the {\em Spitzer} Space Telescope \citep{sst}.  Except
where noted, the assumed cosmological parameters are H$_0 = 70$ km
s$^{-1}$ Mpc$^{-1}$, $\Omega_m = 0.3$ and $\Lambda = 0.7$.

\section{Observations}

\subsection{Sample Definition}

\begin{figure}
\plotone{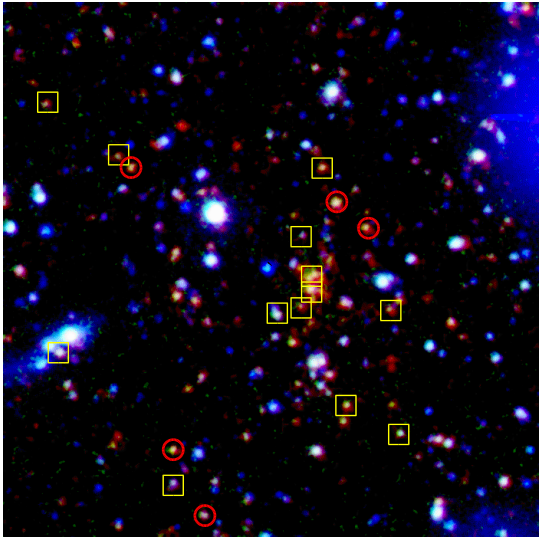}

\caption{$I,K,4.5~\mu$m false-color image of ISCS~J143809+341419 covering 3\arcmin~on a
side.  Objects with photometric redshifts at $1.25 < z < 1.55$ are
marked by the yellow boxes.  Photometric redshifts were not calculated for some of the fainter objects 
in the 4.5$\mu$m catalog so some objects visible in the image which are not marked with boxes in fact may  
be in the cluster.  The 5 spectroscopically confirmed
member galaxies are marked by the red circles. North is up and East is to the 
left in the image.}

\label{color}
\end{figure}

\begin{figure}
\plotone{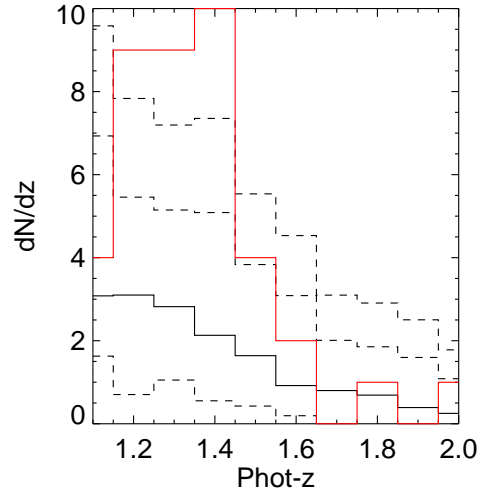}

\caption{Histogram vs redshift showing the photometric redshifts of
galaxies in the IRAC Shallow Survey.  The area covered by the solid red
histogram is $3\arcmin \times 3\arcmin$~ around the $z=1.41$
cluster.  The solid black histogram represents the average of 100
 $3\arcmin \times 3\arcmin$ patches randomly chosen from the entire 
$\sim$9 degree$^2$ IRAC survey in the Bo\"otes field. The dashed histograms 
represent the 1 and 2 $\sigma$ uncertainties on the average value histogram.}

\label{overdensity}
\end{figure}

To identify galaxy cluster candidates in the Bo\"otes field we made
use of the IRAC Shallow Survey \citep{ISS}, the FLAMEX Survey, which
provides $JK_s$ imaging down to $K_s = 19.2$ (R.\ Elston et al.\ in
preparation), and the NOAO Deep Wide-Field Survey (NDWFS), which
provides deep $B_wRI$ imaging \citep{ndwfs}.  A catalog was defined
using the 4.5$\mu$m data from the IRAC survey.  Photometric redshifts
were estimated from matched photometry in the $B_wRIJK$ and the IRAC
3.6 and 4.5 $\mu$m bands.  Then a wavelet search was carried out using
the photometric redshift probability distributions to detect
structures on physical scales of $200 < r < 800$ kpc.  Bootstrap
simulations were then used to quantify the significance of each
detection.  The details of the sample selection and identification of
cluster candidates are given in P.\ Eisenhardt et al.\ (in
preparation), and of the photometric redshift process in M.\ Brodwin et
al.\ (in preparation).  We report here on the candidate cluster
ISCS~J143809+341419, shown in Figure~\ref{color}.  The prefix ISCS
stands for the IRAC Shallow survey Cluster Search.  The cluster
overdensity relative to that of similar size areas in the entire Bo\"otes field in the IRAC
Shallow Survey is shown in Figure~\ref{overdensity}.

\subsection{Keck Spectroscopy}

In order to make sure that the candidate cluster is not a chance
alignment along the line of sight, we obtained optical spectroscopy of
objects in ISCS~J143809+341419.  Objects were selected based on their
photometric redshifts.  The $I$-band magnitudes of the primary galaxy
targets were $22.0 < I < 24.3$.  We prepared a slitmask including slitlets
for 14 objects with photometric redshifts $1.25 < z_p < 1.55$ within 3
arcmin of the nominal cluster center.  The range in $z_p$ was chosen
to cover the nominal photometric redshift of the cluster $\pm 1
\sigma$.  Two other objects in the 4.5$\mu$m catalog were included
so as to fill out the mask.  The slits had widths of 1.3 arcsec and
minimum lengths of 10 arcsec.  The slitmask was used with the Low
Resolution Imaging Spectrograph \citep{LRIS} on the 10~m
Keck II telescope on UT 2005 June 04 to obtain deep spectroscopy.  On
the red side, we used the 400 l mm$^{-1}$ grating, which is blazed at
8500 \AA, to cover a nominal wavelength range of 6000 to 9800 \AA,
depending on the position of a slit in the mask.  The dispersion of
$\sim$1.8 \AA~pixel$^{-1}$ resulted in a spectral resolution of $R
\sim 900$.  On the blue side we used the 400 line grism which is
blazed at 3400 \AA~ and provides coverage from the atmospheric cutoff
up to $\sim$5800 \AA~ where the dichroic splits the light between the
two sides of LRIS.  We obtained 7 $\times$ 1800~s exposures with this
setup in photometric conditions with 0.9 arcsec seeing.  Objects were
shifted along the long axis of the slits between exposures to enable
better sky subtraction and fringe correction.  The observations were
carried out with the slitlets aligned close to the parallactic angle.

The slitmask data were separated into individual spectra and then
reduced using standard longslit techniques.  A relative flux
calibration was obtained from longslit observations of the standard
stars HZ 44 and Wolf 1346.  One--dimensional spectra were extracted
from the sum of all the reduced data for each of the 16 slitlets for
both the red and blue sides.  For the targets in the cluster, only the
red side data are useful and the blue side data will not be discussed
further.

\section{Results}

\subsection{Optical Spectroscopy}

\begin{figure}
\epsscale{1.0}
\plotone{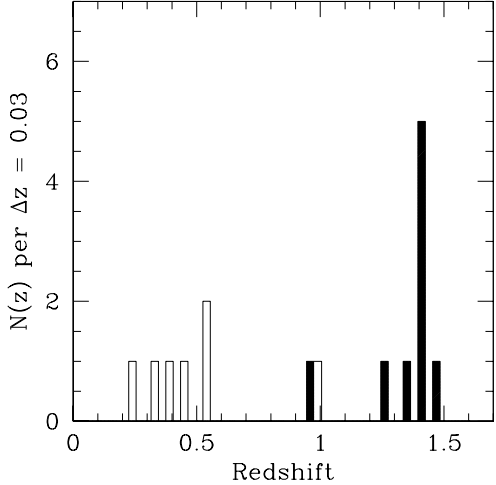}

\caption{Histogram of all redshifts
determined in the Keck/LRIS slitmask observation. The shaded objects are those with photometric 
redshifts $1.25 < z_p < 1.55$.} 

\label{zhist}
\epsscale{1.0}
\end{figure}

Spectra were obtained for 9 of the 14 photometrically-selected
galaxies with quality sufficient for determining the redshifts; the
other 5 were deemed too faint for reliable identification of spectral
features.  Only one of these 9 objects has a redshift outside of the
photometric redshift range.  Four objects have redshifts
within the range $1.4147 < z_s < 1.4172$ so are considered to be
members of a cluster.  A fifth object has $z_s = 1.4028$ which is
somewhat lower than the redshifts of the other four but still likely
to be a cluster member.  A
histogram of all redshifts determined from the mask (including 5 
serendipitous sources) is shown in Figure~\ref{zhist}.  The
spectroscopy validates our photometric redshifts in the range $1.3 < z
< 1.5$ (M. Brodwin et al.\ in preparation).

\begin{figure}
\epsscale{1.0}
\plotone{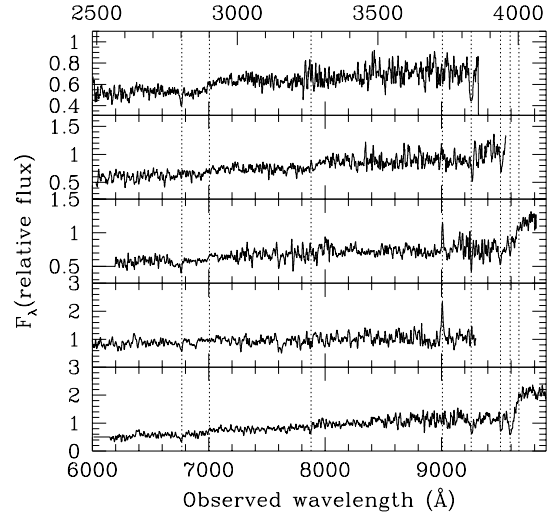}

\caption{Optical spectra of the 5 member galaxies after being smoothed by an 5
pixel boxcar. The following spectral features (though not present in
every object) are marked in increasing wavelength by the dotted lines:
Mg II$\lambda$2800, B2900, B3260, [OII]$\lambda$3727, Mg I$\lambda$3830,
Ca II K and H, and D4000. The rest frame wavelength at $z=1.4166$ is shown along the top. }  

\label{spec}
\epsscale{1.0}
\end{figure}

The spectroscopic members are listed in Table 1.  The table gives the
Vega magnitude [4.5] in the IRAC $4.5~\mu$m band, the photometric and
spectroscopic redshifts, and the template number of the best-fit as
determined by the photometric redshift estimator.  These template
numbers are from \citet{thesis} based on the spectral templates of
\citet{CWW}. The approximate correspondence to
galaxy type spectra is 0 for elliptical, 7 for an Sbc, and 13 for an
Scd. In line with the template fitting, the optical spectra of most of
the members show absorption lines (Ca II H+K, $\lambda 3830$, MgI
$\lambda 2852$ and MgII $\lambda 2800$) and spectral breaks (D4000,
B3260, B2900) similar to those of present--epoch ellipticals.  The
spectra of the member galaxies are shown in
Figure~\ref{spec}.  Only one of the 5 members,IRAC~J143811+341256, has a
prominent emission line; the photometry of this same object was fit best by a
relatively late-type template.  To better determine the redshifts of
the 4 early-type cluster members, we used the Fourier
quotient technique as implemented in the FXCOR task of IRAF.  The red
members were cross--correlated with an early-type galaxy template
\citep{templates}.  The resulting redshifts, along with their
formal uncertainties, are listed in Table 1.  We adopt a nominal
center for the cluster of $\alpha = 14^h 38^m 09\fs4$, $\delta
= +34\arcdeg 14\arcmin 19\arcsec$ (J2000), the centroid of the wavelet
detection, and a nominal redshift $z = 1.413$.  Given the small number
of members it is premature to estimate the velocity dispersion of the
galaxies.

\begin{figure}
\epsscale{1.0}
\plotone{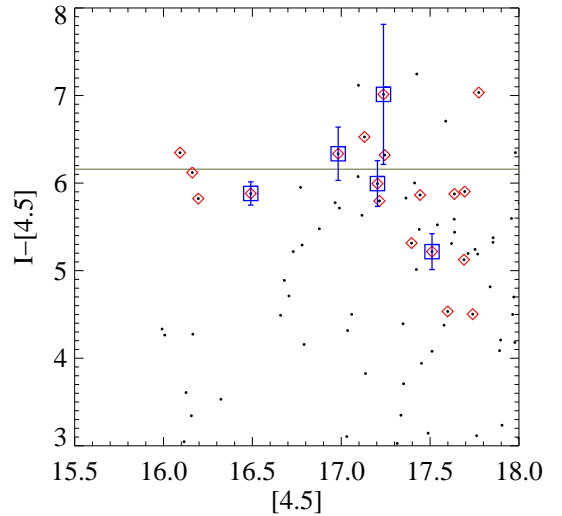}

\caption{Color--magnitude diagram of the cluster in $I - [4.5]$ vs
$[4.5]$.  The magnitudes are in the Vega system and [4.5] is the
magnitude in the $4.5~\mu$m band.  The spectroscopically identified
galaxies at $z=1.41$ are marked with blue squares, and the
photometrically selected galaxies at $1.25 < z < 1.55$ by red
diamonds.  The green horizontal line represents a model prediction at
L$^*$ for a passively evolving stellar population formed in a single
100 Myr burst beginning at $z_f = 3.0$, as described in the text.  The
errors in the photometry are shown only for the spectroscopic member
galaxies. The photometry represents 5\arcsec~corrected aperture
magnitudes. The field size is 3\arcmin $\times$ 3\arcmin.}

\label{cmd} 
\epsscale{1.0} 
\end{figure}

\subsection{Optical and Infrared Photometry}

A color--magnitude diagram for all objects in a 3\arcmin~area at the
cluster are shown in Figure~\ref{cmd}.  The objects with photometric
redshifts in the range $1.25 < z < 1.55$ are shown by the red diamonds,
and those with spectroscopic redshifts $1.40 < z < 1.42$ are shown by
the blue squares.  The scatter seen in the colors of the member
galaxies is dominated by the photometric uncertainties in the NDWFS photometry of $I \gtrsim 23$ galaxies. 

Also plotted in Figure~\ref{cmd} is an estimate of the expected color
for L$^*$ early--type galaxies. Using a simple passive evolution model
calculated from the GISSEL of \citet{BC} and our assumed
cosmology, we calculated the expected colors for a single 100 Myr
burst stellar population formed at $z_f = 3.0$.  As can be seen in
Figure~\ref{cmd}, the colors of the brighter member galaxies are consistent with 
those exected for early-type galaxies whose
stars formed at $z_f = 3.0$.   The three brightest objects on the color-magnitude relation have not 
yet been observed spectroscopically.

\section{Discussion}

The observations presented here provide strong evidence supporting the
identification of ISCS~J143809+341419 as a galaxy cluster at $z = 1.41$.
Photometric redshifts, demonstrated to be accurate by the optical
spectroscopy in the relevant redshift range, indicate a significant
overdensity of galaxies within a region of radius $r = 300$ kpc at $z \sim 1.4$.
Our optical spectroscopy shows that 5 of these galaxies are at the
same redshift.  Further examination of the properties of the cluster,
such as its mass and the evolutionary state of its galaxies, is
deferred until spectroscopy can confirm more members and deeper
multiband imaging can be obtained.

It is noteworthy that ISCS~J143809+341419 was detected using imaging taken
by IRAC with exposures lasting only 90~s per pointing over the $\sim$9
degree$^2$ area in the Bo\"otes field of the NDWFS.  As implied by the
identification of another $z \sim 1.4$ cluster which relied on a short
20~ks exposure in the XMM archive \citep{mullis}, current and
planned surveys are better off covering larger areas rather than going
deep if the goal is to find $z > 1$ galaxy clusters. To date three
other $z > 1$ IRAC-selected clusters have been spectroscopically
confirmed in the Bo\"otes field (P.\ Eisenhardt et al.\ in preparation, M.\ Brodwin et
al.\ in preparation, and R.\ Elston et al.\ in preparation).  With the combination of the
search described here in the Bo\"otes field, serendipitous searches
for clusters in the XMM archival data \citep{XCS,mullis}, and
the on-going wide-area surveys such as SWIRE \citep{swire},
the RCS \citep{RCS}, and the CFHT-Legacy Survey, the time finally is
ripe for the identification of large samples of $z > 1$ clusters.  The construction of 
such samples will pave the way towards a better understanding of the origin 
of early-type galaxies.

\acknowledgments

This work is based on observations made with the Spitzer Space
Telescope, which is operated by the Jet Propulsion Laboratory,
California Institute of Technology.  The W.\ M.\ Keck Observatory is a
scientific partnership between the University of California and the
California Institute of Technology, made possible by a generous gift
of the W.\ M.\ Keck Foundation. The authors wish to recognize and
acknowledge the very significant cultural role and reverence that the
summit of Mauna Kea has always had within the indigenous Hawaiian
community; we are most fortunate to have the opportunity to conduct
observations from this mountain.  The Flamingos Extragalactic Survey
and the NOAO Deep Wide-Field Survey would not have been possible
without support from NOAO, which is operated by the Association of
Universities for Research in Astronomy, Inc., under a cooperative
agreement with the National Science Foundation (NSF).  We thank the
observing teams at the University of Florida and at NOAO for
supporting these two surveys. A.\ H.\ G.\ acknowledges support from
the NSF AAPF under award AST-0407085 and from an NSF SGER under award
AST-0436681. We thank the referee for comments leading to an improved version of 
the paper.

\begin{deluxetable}{ccccccc}
\small
\tablenum{1}
\tablecaption{Summary of Spectroscopic Cluster Members}
\tablehead{
\colhead{ID} & 
\colhead{R.A.\tablenotemark{a}} & 
\colhead{Dec.\tablenotemark{a}} & 
\colhead{[4.5]} & 
\colhead{phot-z} & 
\colhead{template} & 
\colhead{spec-z} 
}
\startdata
IRAC~J143806+341433 & 14:38:06.97 & 34:14:33.8 & 16.68 & 1.39 & 1	& 1.4153  \\
IRAC~J143807+341441 & 14:38:07.80 & 34:14:41.6 & 16.18 & 1.30 & 3	& 1.4166  \\  
IRAC~J143813+341452 & 14:38:13.36 & 34:14:52.9 & 16.93 & 1.48 & 0	& 1.4028  \\
IRAC~J143812+341318 & 14:38:12.10 & 34:13:18.2 & 16.90 & 1.35 & 3	& 1.4147  \\
IRAC~J143811+341256 & 14:38:11.25 & 34:12:56.2 & 17.20 & 1.39 & 11	& 1.4172  \\
\enddata
\tablenotetext{a}{Coordinates are J2000.}
\end{deluxetable}

\end{document}